\documentclass[twocolumn, revtex4,iop, twocolappendix]{oja}
\usepackage{newtxtext,newtxmath}

\usepackage[T1]{fontenc}

\DeclareRobustCommand{\VAN}[3]{#2}
\let\VANthebibliography\thebibliography
\def\thebibliography{\DeclareRobustCommand{\VAN}[3]{##3}\VANthebibliography}

\usepackage{graphicx}	
\usepackage{amsmath}	
\usepackage{amssymb}	
\usepackage{booktabs}

\usepackage[colorlinks,linkcolor=blue,citecolor=blue,urlcolor=blue ]{hyperref}
\usepackage{cleveref}

\begin{document}

\title[Host dispersion as feedback probe]
{Galaxy dispersion measured by Fast Radio Bursts as a probe of baryonic feedback models}
\author{Alexander Theis}
\email{alexander.theis@campus.lmu.de}
\affiliation{Universitäts-Sternwarte, Fakultät für Physik, Ludwig-Maximilians Universität München,
Scheinerstraße 1, D-81679 München, Germany}

\author{Steffen Hagstotz}
\affiliation{Universitäts-Sternwarte, Fakultät für Physik, Ludwig-Maximilians Universität München, 
Scheinerstraße 1, D-81679 München, Germany and\\
Excellence Cluster ORIGINS, Boltzmannstraße 2, D-85748 Garching, Germany}
\email{steffen.hagstotz@lmu.de}

\author{Robert Reischke}
\affiliation{Ruhr University Bochum, Faculty of Physics and Astronomy, Astronomical Institute (AIRUB)  German Centre for Cosmological Lensing, 44780 Bochum, Germany and\\
Argelander-Institut für Astronomie, Universität Bonn, Auf dem Hügel 71, D-53121 Bonn, Germany}

\author{Jochen Weller}
\affiliation{Universitäts-Sternwarte, Fakultät für Physik, Ludwig-Maximilians Universität München, 
Scheinerstraße 1, D-81679 München, Germany\\
Excellence Cluster ORIGINS, Boltzmannstraße 2, D-85748 Garching, Germany and\\
Max Planck Institute for Extraterrestrial Physics, Giessenbachstrasse 1, 85748 Garching, Germany}


\begin{abstract}
Fast Radio Bursts (FRBs) are a sensitive probe of the electron distribution in both the large-scale structure and their host galaxies through the dispersion measure (DM) of the radio pulse. Baryonic feedback models are crucial for modelling small scales for ongoing cosmological surveys that are expected to change the electron distribution in galaxies in a way that can be probed by FRB observations. In this paper, we explore the impact of baryonic feedback on FRB hosts using numerical simulations and make a detailed study of the host galaxy dispersion as a function of redshift, galaxy type, feedback model and how these properties vary in independent simulation codes.
We find that the host galaxy dispersion varies dramatically between different implementations of baryonic feedback, allowing FRBs with host identification to be a valuable probe of feedback physics and thus provide necessary priors for upcoming analysis of the statistical properties of the large-scale structure.

We further find that any dependency on the exact location of events within the halo is small. While there exists an evolution of the dispersion measure with redshift and halo mass, it is largely driven by varying star formation rates of the halo. Spectral information from FRB hosts can therefore be used to put priors on the host galaxy dispersion measure, and FRBs can be used to distinguish between competing models of baryonic feedback in future studies.
\end{abstract}


\section{Introduction}

Fast radio bursts (FRBs) are very short (ms) transients observed in frequencies from $\sim 100$ MHz up to a few GHz. While overwhelmingly originating from extragalactic sources, their actual progenitors remain unknown, resulting in a variety of proposed models, including mergers \citep{Liu_2016} and magnetars \citep{Thornton_2013}. The latter, closely associated with some FRBs \citep{Bochenek_2020}, are particularly promising candidates. FRBs can serve as promising probes of astrophysics and cosmology, with an increasing number of events detected globally by telescopes such as the Canadian Hydrogen Intensity Mapping Experiment \citep[CHIME, e.g.][]{rafiei-ravandi_characterizing_2020}. As the radio signal traverses ionized and magnetized media like the intergalactic medium (IGM), it undergoes dispersion, providing valuable information encoded in the dispersion measure \citep[$\mathrm{DM}$, see for example][]{thornton_population_2013,zhou_fast_2014,petroff_real-time_2015,connor_non-cosmological_2016,champion_five_2016,chatterjee_direct_2017,Macquart:2020lln}. The $\mathrm{DM}(z)$ holds significance in astrophysics and cosmology, contributing to addressing current challenges like the Hubble tension \citep{Macquart:2020lln,Hagstotz_2022,Wu:2020jmx, James_2022} or constraining the equivalence principle \citep{reischke_consistent_2023} via the cosmological covariance between the FRBs \citep{reischke_cosmological_2023}. 
 Cosmology with FRBs can, potentially, be based on two pillars. $(i)$ The aforementioned redshift dependence, the $\mathrm{DM}(z)$ relation, probes the background cosmology and obtains its uncertainty from cosmic variance (recently \citet{flimflam_2024} discussed the possibility of reducing cosmic variance via density field reconstruction using external galaxy surveys) and the scatter in the host contribution.
$(ii)$ Statistical properties of the $\mathrm{DM}$ via e.g. two-point functions as discussed in \citet{masui_dispersion_2015,shirasaki_large-scale_2017,Reischke:2020cgd,Reischke:2021euf,rafiei-ravandi_characterizing_2020,bhattacharya_fast_2020,alonso_linear_2021,takahashi_statistical_2021,reischke_feedback_2023}.

\label{sec:sims}
 \begin{figure*}
    \centering
    \includegraphics[width=0.7\textwidth]{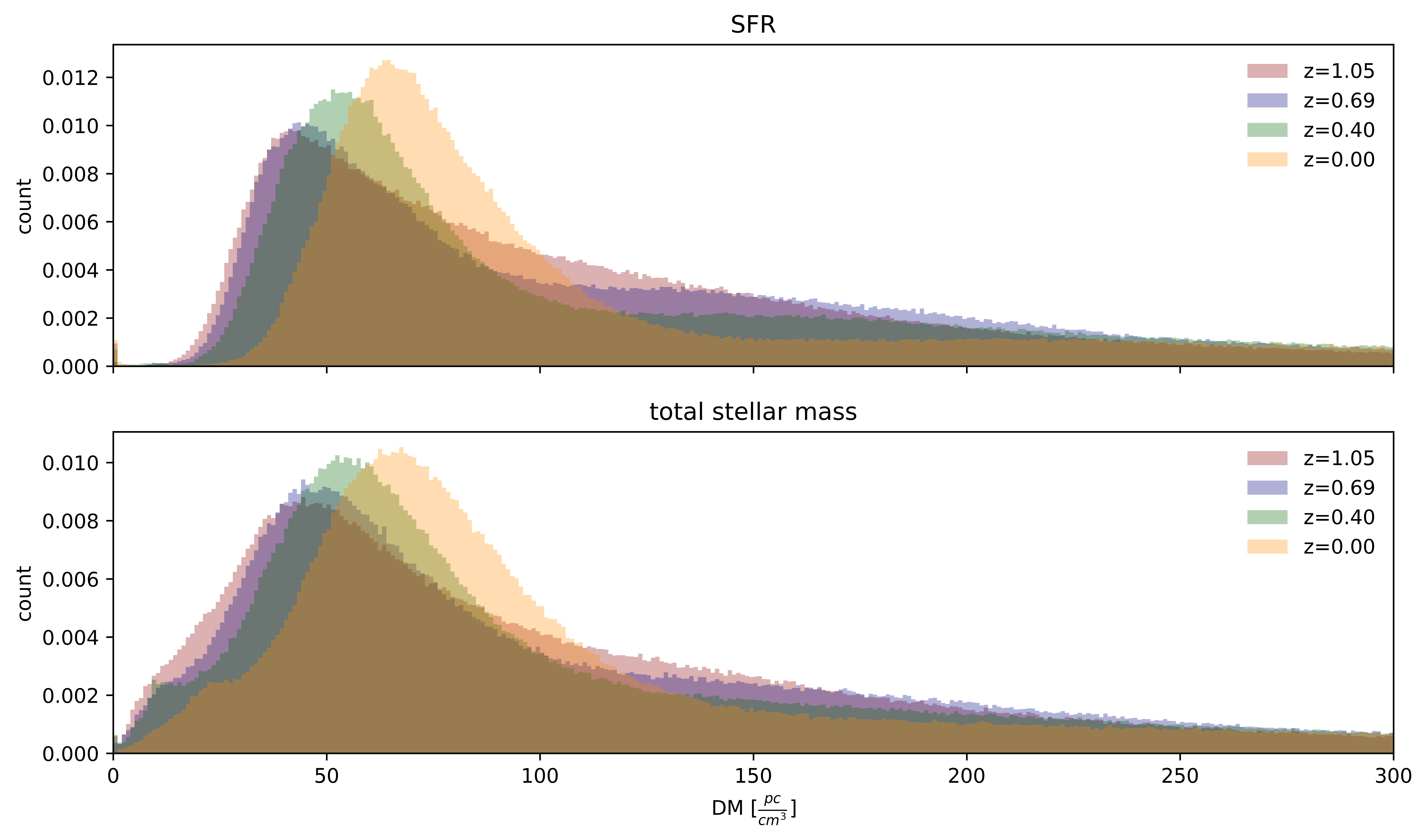}
    \caption[Redshift dependency of $\mathrm{DM}_{\mathrm{host}}$]{Normalised histogram of the host galaxy contribution to the DM illustrating its redshift dependency for the IllustrisTNG suite. Note that we always show rest-frame quantities.   
   The upper and lower panels show the difference in placing the FRBs either tracing the star formation rate (SFR) or the total stellar mass respectively. }
    \label{fig:redshift_dependency}
    \end{figure*}

However, localizing FRBs to their host galaxies remains challenging, limiting the available data for cosmological applications and despite its potential, FRB-based cosmological probes face challenges, including uncertainties in the individual contributors to the observed $\mathrm{DM}$, namely Milky Way, IGM, and host contributions. While models exist for the first two, the host contribution $\mathrm{DM}_{\mathrm{host}}$ lacks a well-defined description. Currently, it is pretty standard to assume a Gaussian or log-normal distribution for the host contribution, whether this is a good approximation, however, needs to be tested, especially in hydrodynamical simulations.

This is the reason for this paper to propose a detailed model of the host galaxy dispersion contribution as a function of redshift, feedback and galaxy properties based on the Cosmology and Astrophysics with MachinE 
Learning Simulations \citep[CAMELS,][]{Villaescusa_Navarro_2021}. The simulation suite offers the opportunity to measure and compare FRB host properties in various settings.

The paper is structured as follows: we go through the basics of FRB cosmology in Section \ref{sec:frb_intro} and discuss the simulations briefly in Section \ref{sec:sims}. In Section \ref{sec:sims_host} we discuss the host contribution of the FRBs as a function of different parameters and provide a fitting function. We then conclude in Section \ref{sec:conclusion}.

\section{Fast Radio Burst Dispersion}
\label{sec:frb_intro}

The radio pulse emitted by the FRB source undergoes dispersion as it traverses the ionized intergalactic medium. This dispersion leads to a characteristic time delay between the arrival times of different pulse frequencies, denoted as $\nu$, given by:
\begin{equation}
\label{eq:DM_def}
    \Delta t = \mathrm{DM}(\mathbf{x},z) \nu^{-2}\,,
\end{equation}
\noindent This relationship defines the inferred dispersion measure (DM) of an FRB at the sky position $\mathbf{x}$ and redshift $z$, arising from free electrons along the line of sight. The total dispersion measure can then be written as
\begin{equation}
    \mathrm{DM}_\mathrm{tot} = \int \frac{n_e(l)}{1+z(l)} \mathrm{d} {l} \, ,
\end{equation}
for any line of sight parameterised by $l$. Note that because of the frequency dependence in Eq.~\ref{eq:DM_def}, the effect of the electron density at any point is weighted by $1/(1+z)$ as the pulse gets redshifted \citep{Macquart:2020lln}.

Electrons may be associated with either the host halo, the Milky Way and its halo or the large-scale structure (LSS). Therefore, we can write the total contribution as the following:
\begin{equation}
    \mathrm{DM}_{\mathrm{tot}}(\mathbf{x}, z) = \mathrm{DM}_{\mathrm{host}}(z) + \mathrm{DM}_{\mathrm{MW}}(\mathbf{x}) + \mathrm{DM}_{\mathrm{LSS}}(z,\mathbf{x})\,,
\end{equation}
\noindent Here, $\mathrm{DM}_{\mathrm{MW}}(\mathbf{x})$ defines the contribution from the Milky Way, which can be described by the NE2001 \cite{cordes2003ne2001} or YMW16 \citep{YMW16} models calibrated to the dispersion of pulsars within the galaxy, and an uncertainty of $\sigma_{\mathrm{MW}}$ to account for the model uncertainty and contributions from the extended Milky Way halo. Often, the quantity of cosmological interest is the mean contribution of electrons associated with the large-scale structure $\mathrm{DM}_{\mathrm{LSS}}(z)$,  which can be calculated from the cosmological background densities via \citep{Macquart:2020lln}:
\begin{equation}\label{eq:Macquart_relation}
    \langle \mathrm{DM}_{\mathrm{LSS}} \rangle (z) = \frac{3 H_0 \Omega_\mathrm{b}}{8 \pi G m_\mathrm{p}}  \chi_\mathrm{e} f_{\mathrm{IGM}} \int_0^z \frac{(1+z')}{E(z')} \mathrm{d}z' \; .
\end{equation}
with redshift $z$, the electron fraction $\chi_\mathrm{e} \approx 1 - Y_\mathrm{He} = 0.88$, which is calculated from the cosmic helium abundance \citep{Aver:2015iza}, the proton mass $m_\mathrm{p}$, the fraction of electrons in the IGM $f_{\mathrm{IGM}} \approx 0.84$, which can be estimated from the electrons bound in compact objects and dense gas \citep{Madau:2014bja}, and the dimensionless Hubble constant $E(z) = H(z) / H_0$. Lastly, the final contribution from the FRB's host is the DM picked up while the burst travels through the electron distribution of its origin galaxy. For this contribution, the redshift dependence is again made explicit because the restframe DM is again scaled with $(1+z)^{-1}$:
\begin{equation}
\label{eq:DM_host}
    \mathrm{DM}_{\mathrm{host}}(z) = \frac{\mathrm{DM}_{\mathrm{host,0}}}{1+z}\,,
\end{equation}
and from now on we will always refer to the rest-frame host dispersion and drop the index $\mathrm{DM}_\mathrm{host,0}$. Since the redshift evolution of \cref{eq:Macquart_relation} and \cref{eq:DM_host} differs, measurements of FRBs can in principle distinguish the amplitudes of both terms independently as long as observations span a sufficient range of redshifts. The only exception would be an intrinsic evolution of $\mathrm{DM}_\mathrm{host}$ that mimics the one of \cref{eq:Macquart_relation} coincidentally, which we investigate in Section \ref{results}. With limited data available, the host dispersion is a crucial input for interpreting the cosmological information contained in \cref{eq:Macquart_relation} and is a relevant nuisance parameter for studies of either the mean $\mathrm{DM}(z)$ relation \citep{Hagstotz_2022, James_2022} or enters as a noise term in studies of the dispersion measure correlations \citep{reischke_feedback_2023}. However, as we discuss in this paper, even the bare host DM, $\mathrm{DM}_{\mathrm{host,0}}$, and its evolution with redshift or halo properties contains information about galaxy evolution, and is strongly affected by baryonic feedback processes.

\begin{table}
\centering
\begin{tabular}{cccccc}
\toprule
\textbf{$\Omega_\mathrm{b}$} & \textbf{$h$} & \textbf{$n_s$} & \textbf{$w_{\Lambda}$} & \textbf{$M_{\nu} [eV]$} & \textbf{$\Omega_k$} \\
\midrule
$0.049$ & $0.6711$ & $0.9624$ & $-1$ & $0.0$ & $0.0$ \\
\bottomrule
\end{tabular}
\caption{Table with the fixed cosmological parameters used for the CAMELS simulations of this analysis where $n_s$ is the spectral index and $M_{\nu}$ represents the sum of neutrino masses \citep{Villaescusa_Navarro_2021}.}
\label{tab:fixed_cosmological_params}
\end{table}

Pinning down this quantity is key, if one wants to use FRBs as a cosmological probe and can itself provide valuable information about the power spectrum at nonlinear scales.

\section{Simulations}
The Cosmology and Astrophysics with MachinE Learning Simulations \citep[CAMELS,][]{Villaescusa_Navarro_2021} project provides the simulation data for predicting the distribution of host halos $\mathrm{DM}_{\mathrm{host}}$. The CAMELS project aims to connect cosmology and astrophysics through numerical simulations and machine learning. It encompasses thousands of simulations with diverse cosmological parameters and underlying astrophysical models. Currently, CAMELS includes 10,680 simulations, of which 5,516 include hydrodynamical models. These simulations track the evolution of $256^3$ dark matter particles and $256^3$ gas particles within a periodic comoving volume of $(25 \mathrm{Mpc} \; \mathrm{h}^{-1})^3$ starting at redshift $z = 127$, providing snapshots from $z = 15$ to $z = 0$. The hydrodynamical simulations in CAMELS are categorized into different suites representing galaxy formation models based on distinct subgrid physics, more specifically IllustrisTNG \citep{weinberger_tng_2017,pillepic_tng_2018}, SIMBA \citep{hopkins_gizmo_2015,dave_2019_simba} and ASTRID \citep{bird_astrid_2022,ni_astrid_2022} which we compare in this paper. In addition, CAMELS provides various simulation sets for each suite, including the one-parameter variation (1P) and the cosmic variance sets (CV) which are the foundation of this paper's work. The 1P sets explore the effects of specific parameter variations, keeping initial conditions constant, whereas the CV sets vary only the initial conditions \citep{Villaescusa_Navarro_2021, Villaescusa_Navarro_2023, ni2023camels}. However, all simulations used in this study share the cosmological parameters shown in table \ref{tab:fixed_cosmological_params}. Apart from those fixed parameters, the utilized version of CAMELS features the variation of six parameters, namely the total matter density parameter $\Omega_m$, the root mean square amplitude of density fluctuations $\sigma_8$ measured within a sphere of radius $8\; \mathrm{Mpc} \; \mathrm{h}^{-1}$ at redshift $z = 0$, and two astrophysical parameters each for both AGN and SN feedback. $A_\mathrm{SN1}$ and $A_\mathrm{SN2}$ control the properties of galactic winds, in particular, their speed and efficiency (e.g. via their mass loading). For the AGN feedback, $A_\mathrm{AGN1}$ and $A_\mathrm{AGN2}$ control the BH accretion speed and efficiency respectively. As discussed in \citet{ni2023camels} we expect ASTRID to have the weakest feedback and therefore a very clustered large-scale baryon distribution. On the other hand, SIMBA and IllustrisTNG will have smoother baryon distributions due to stronger feedback.
The specific and detailed interpretation of those astrophysical parameters in each of the three suites is given in \cite{ni2023camels}. The key argument, however, is that the true feedback model is not known and different implementations of the sub-grid physics might yield very different results which then need to be distinguished by the data.

\section{Methodology}
\label{sec:sims_host}
This part serves the purpose of presenting the methodology of the halo selection as well as details about the host dispersion measure calculations through the placement of mock FRB events in the simulated halos and tracing their dispersion measure.

\subsection{Halo selection and exclusion criteria}

The selection of the halos is based on the CAMELS' SUBFIND catalogue \citep{Springel:01, Dolag_2009}. This catalogue provides crucial information about each halo, including its mass, virial radius, and the number of subhalos\footnote{A comprehensive list of all available quantities can be found in \cite{TNG} or on \url{https://camels.readthedocs.io/en/latest/subfind.html}.}. Initially, the selection process focuses on halos within the mass range of $10^{10} \frac{M_{\odot}}{h} < M_{\mathrm{halo}} < 10^{12.5}\frac{M_{\odot}}{h}$, which is a fiducial estimate for the galaxy scale of interest. Individual halos are then determined by considering all gas particles within a radius of $3 r_{\mathrm{vir,halo}}$ around the halo's centre. This choice aligns with previous works \citep{Mo_2022} and has been validated through an analysis of the radius's impact on $\mathrm{DM}_{\mathrm{host}}$. After the initial halo selection process, we employed additional exclusion criteria for this study. Beyond the mass constraint, halos devoid of any star formation rate, i.e. $\Sigma_i \mathrm{SFR}_i = 0$ with $i$ being an index for gas particles, were excluded to avoid errors during the source placement process. Additionally, halos with fewer than $25$ star particles were excluded to ensure more robust statistical analyses. The remaining halos are mostly constrained to galaxy and group sizes as shown in Appendix \ref{app:halos}, due to the limited box size of CAMELS. This limits the ability of the simulations to study the properties of massive halos and galaxy clusters.

\begin{figure}
    \centering
    \includegraphics[width=0.47\textwidth]{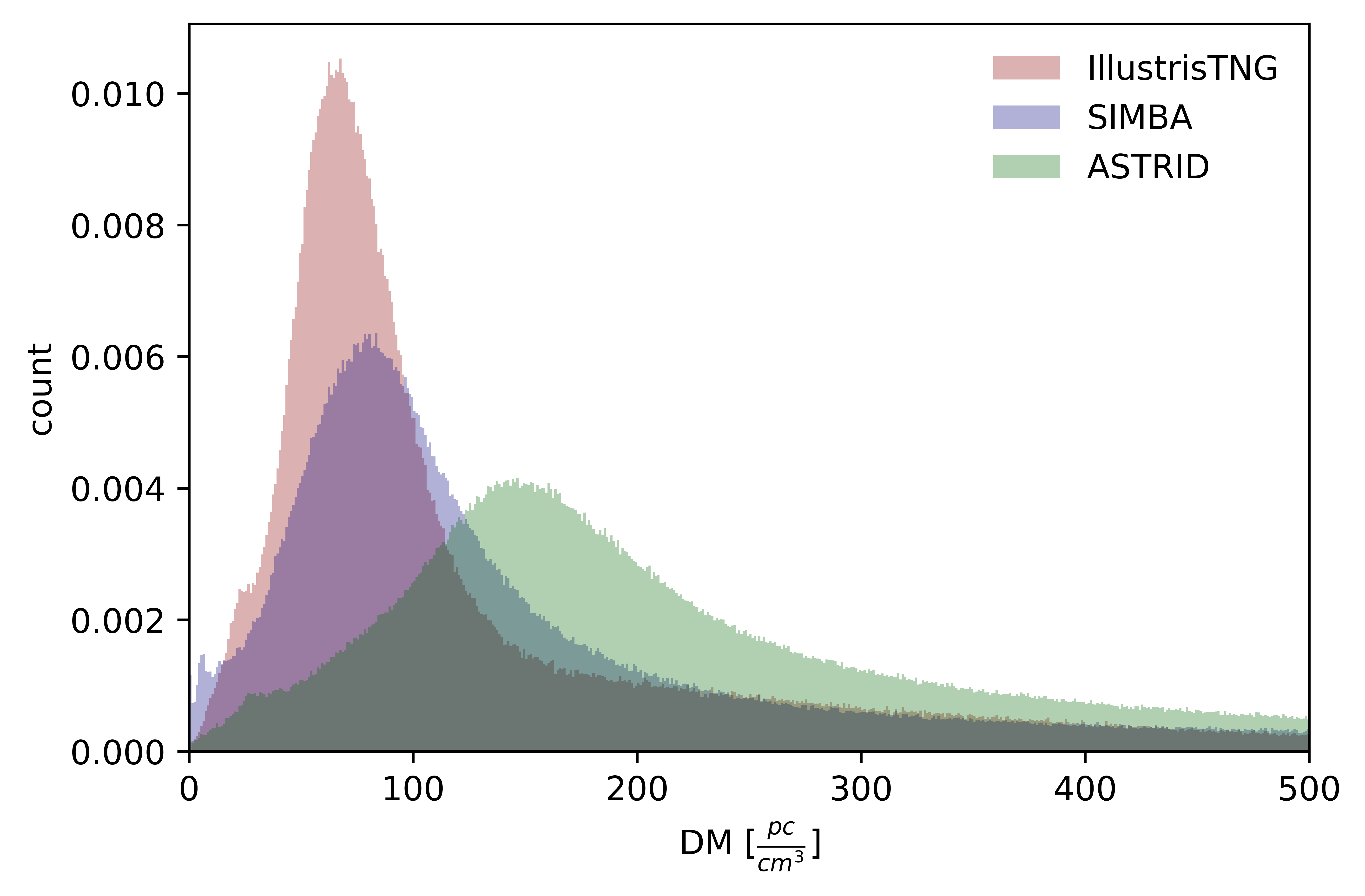}
    \caption[Suite comparison $\mathrm{DM}_{\mathrm{host}}$]{Normalised histogram of the host galaxy contribution to the $\mathrm{DM}_{\mathrm{host}}$ distribution over all halos among the three suites, as discussed in the text. Each histogram utilizes the CV0 set and the snapshot corresponding to $z=0$.}
    \label{fig:suite_comparison}
    \end{figure}
\subsection{Source placement and calculation of DM}

Since the progenitors of FRBs are not known, it is not a priori clear how to place events within the selected halos before calculating line-of-sight dispersion measures. The main competing results are based on observations of short and bright radio emission from a galactic magnetar \citep{Bochenek_2020}, which would favour an FRB population tracing young stars or the star formation rate. On the other hand, the localisation of another FRB to a globular cluster \citep{Kirsten:2021llv} makes it seem plausible that FRBs follow just the total stellar mass distribution if they are associated with another stellar phenomenon.

 \begin{figure}
    \centering
    \includegraphics[width=0.47\textwidth]{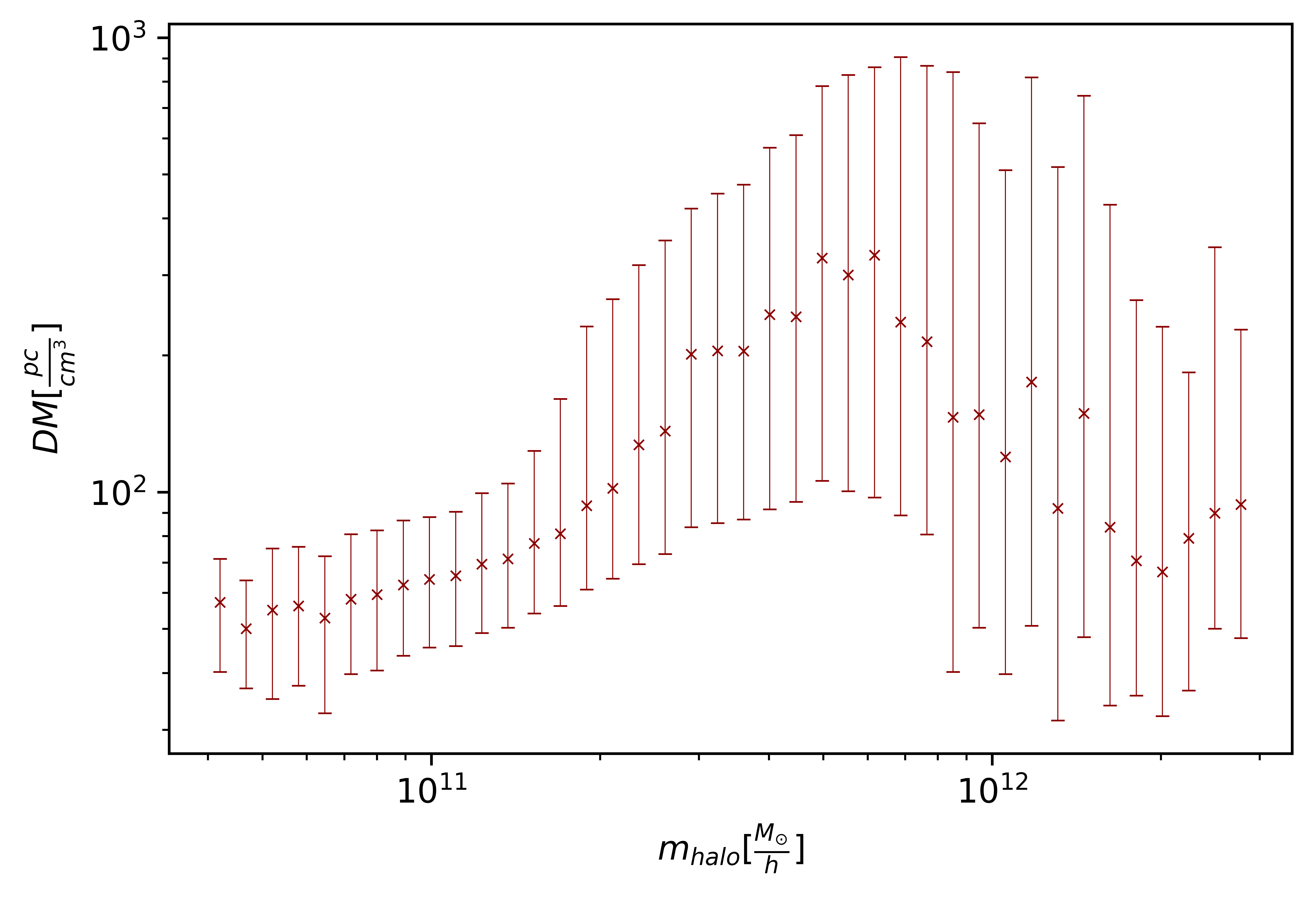}
    \caption[halo mass vs $\mathrm{DM}_{\mathrm{host}}$]{This graph shows the dependency of $\mathrm{DM}_{\mathrm{host}}$ on the halo mass for the IllustrisTNG suite for the joined datasets (red) of all snapshots with $z \in [0.00, 1.05]$. The error bars show the statistical scatter of halos in the given mass bin. The error bars in this plot only reflect the variance between the lines of sight through the existing halos, so they do not reflect the limited number of halos available. Note that halos above $10^{12} \: M_\odot $ are very rare due to the limited box sizes of the simulations (see Appendix \ref{app:halos}).}
    \label{fig:m_halo_vs_DM}
    \end{figure}

To account for this uncertainty, we employ two approaches for placing Fast Radio Bursts (FRBs). The first option assumes that FRBs are associated with magnetars or young stars, and as a proxy we base the placement on the star formation rate ($\mathrm{SFR}$) at each particle position, determining the placement probability as:
\begin{equation*}
    p_{\mathrm{SFR}, i} = \frac{\mathrm{SFR}_i}{\Sigma_j \mathrm{SFR}_j}\,.
\end{equation*}
\noindent This approach highlights the likelihood of FRB sources originating from regions with higher star formation rates. The second method considers the total stellar mass (TSM), with the placement probability calculated as:
\begin{equation*}
    p_{\mathrm{stellar, i}} = \frac{m_{s,i}}{\Sigma_j m_{s,j}}\;.
\end{equation*}
\noindent Here, $m_{s,i}$ represents the star particle mass at position $i$. These placement modes, used previously in \cite{Mo_2022}, help discern whether FRB sources predominantly come from younger source models ($\mathrm{SFR}$) or older ones ($\mathrm{TSM}$). After the events are placed, we sample 20 random sight lines for each FRB uniformly over angular directions.

After completing the aforementioned steps, the final task involves calculating the dispersion measure ($\mathrm{DM}$) along each line of sight ($\mathrm{l.o.s.}$). To achieve this, numerical line integration is employed:
\begin{equation*}
    \mathrm{DM}_{\mathrm{l.o.s.}} = \sum_{i} \Delta l_i n_{e,i}\,.
\end{equation*}
Recall again that all quantities are expressed in the restframe so that the factor $a  = (1+z)^{-1}$ always drops out.

Each path is divided into 1000 segments of length $\Delta l_i$, irrespective of its overall dimension. At each step $i$, the free electron density $n_{e,i}$ at the closest particle position is determined using a KDTree implementation. This process is repeated for every line of sight from any source position, and the results are aggregated to obtain the $\mathrm{DM}_{\mathrm{host}}$ distribution for individual halos. We perform this integral up to $3 r_\mathrm{vir}$ away from the halo centre and explicitly check that variations in the integration boundary up to a few virial radii do not affect the results discussed here. The resulting halo dispersion measure distributions are subsequently utilized for the final analyses in this work.

  \begin{figure}
    \centering
    \includegraphics[width=0.47\textwidth]{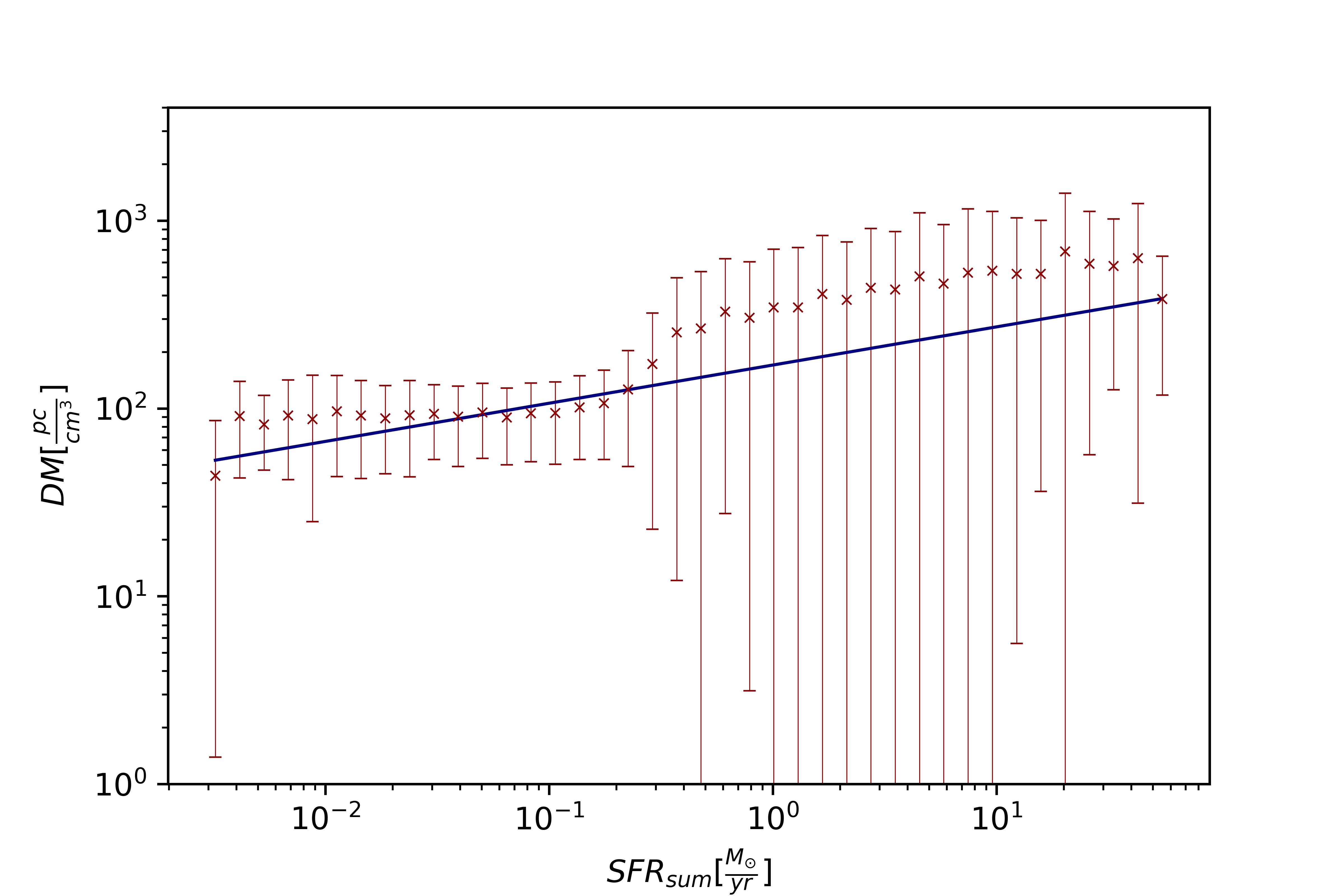}
    \caption[$\mathrm{SFR}$ vs $\mathrm{DM}_{\mathrm{host}}$]{This graph shows the dependency of $\mathrm{DM}_{\mathrm{host}}$ on the $\mathrm{SFR}$ for the IllustrisTNG suite for the joined datasets (red) of all snapshots with $z \in [0.00, 1.05]$ as well as our best-fit model prediction (blue).}
    \label{fig:SFR_vs_DM}
    \end{figure}

\section{Results}\label{results}

\subsection{Source placement dependency}

Our analysis revealed that, regardless of the simulation suite, redshift, or mass range considered, the mean value of $\mathrm{DM}_{\mathrm{host}}$ remains relatively consistent for both placement modes—following either the $\mathrm{SFR}$ or the total stellar mass as seen in \Cref{fig:redshift_dependency}. However, a discernible trend emerged, indicating that the total stellar mass placement mode tends to result in a slightly broader distribution compared to the one following the $\mathrm{SFR}$. This trend might find an explanation in the evolution of stars. As stellar populations migrate towards the centre of the halo over their lifespan, one would anticipate increased variability in the resulting $\mathrm{DM}_{\mathrm{host}}$ distribution when the source placement follows the total stellar mass. Nonetheless, we conclude that the origin of FRBs depending on the local environment of the host does not significantly alter the host contribution.

Furthermore, we can see that the host contribution increases with a function of redshift. While the SFR rate peaks around $z\approx 2$, the growth of galaxies due to hierarchical structure growth leads to more free electrons in hosts due to more massive halos on average. The CAMELS simulations make this difficult to investigate, since beyond $z \sim 1$ the number of halos is quite small due to the limited box size.

We want to emphasize that the intrinsic redshift evolution of $\mathrm{DM}_\mathrm{host}$ is only mild, and differs from the $\mathrm{DM}_\mathrm{LSS}$ term in \cref{eq:Macquart_relation} substantially. We therefore expect that FRBs are indeed able to measure the terms $\mathrm{DM}_\mathrm{LSS}$ and $\mathrm{DM}_\mathrm{host}$ independently.

Overall, these smaller differences are overwhelmed by other effects discussed in the next section. It appears unlikely that the host DM can yield strong constraints on the progenitor of FRBs directly. However, population studies based on large samples split between galaxy types could still reveal a statistical preference for possible FRB mechanisms. We leave a detailed study for future work.

\subsection{IllustrisTNG vs ASTRID vs SIMBA}
Following our analysis of the suites provided by CAMELS, we found a median $\mathrm{DM}_{\mathrm{host}}$ including the $16^{\mathrm{th}}$ and $84^{\mathrm{th}}$ percentile at $z=0$ of $90^{+211}_{-47} \; \mathrm{pc} \; \mathrm{cm}^{-3}$, $131^{+570}_{-68} \; \mathrm{pc} \; \mathrm{cm}^{-3}$ and $209^{+331}_{-95} \; \mathrm{pc} \; \mathrm{cm}^{-3}$ for the IllustrisTNG, SIMBA and ASTRID simulation, respectively. This can also be seen in \Cref{fig:suite_comparison} which illustrates the total $\mathrm{DM}_{\mathrm{host}}$ distributions for the three suites at $z=0$. However, due to the apparent heavy long tail function in the $\mathrm{DM}_{\mathrm{host}}$ distributions, we also computed the maximum probability (MP) value of $\mathrm{DM}_{\mathrm{host}}$ for a more refined understanding. 
Consequently, the analysis yielded $\mathrm{DM}_{\mathrm{host}}$ values of $67 \; \mathrm{pc} \; \mathrm{cm}^{-3}$ for IllustrisTNG, $83 \; \mathrm{pc} \; \mathrm{cm}^{-3}$ for SIMBA, and $147 \; \mathrm{pc} \; \mathrm{cm}^{-3}$ for ASTRID. Notably, ASTRID exhibited larger $\mathrm{DM}_{\mathrm{host}}$ values, potentially attributed to overall weaker feedback \citep{ni2023camels}, resulting in a more condensed halo and electron density ($n_e$) structure. The findings align with earlier studies on data, such as \cite{James_2022}, who reported a median $\mathrm{DM}_{\mathrm{host}}$ of $186_{-48}^{+59} \; \mathrm{pc} \; \mathrm{cm}^{-3}$ measured from localised FRBs. While IllustrisTNG peaks at a much lower $\mathrm{DM}_{\mathrm{host}}$ than ASTRID, the long tail of the distribution makes it still consistent with those observations, allowing for large outliers.
Thus, our results emphasize that systematic differences among the simulation suites, especially between ASTRID and IllustrisTNG or SIMBA, are more significant than variations due to redshift, source placement, or feedback dependencies (see Section \ref{sec:feedbackdep}). To detect the difference in the host galaxy contribution typically a few hundred FRBs are required, yielding a $<10\%$ measurement of the $\mathrm{DM}_{\mathrm{host}}$ contribution \citep{Hagstotz_2022}. 

These results show that the host contribution can be used to distinguish baryonic feedback models from numerical simulations with vastly different implementations and resulting baryon and matter distributions. Again, our results reflect the expectation that ASTRID has much weaker feedback than IllustrisTNG and therefore halos are able to keep more baryons. Using FRBs to inform feedback modelling and therefore the shape of the nonlinear cosmological power spectrum can thus improve constraints on cosmological parameters from other cosmological probes.

    \begin{figure*}
    \centering
    \includegraphics[width=0.9\textwidth]{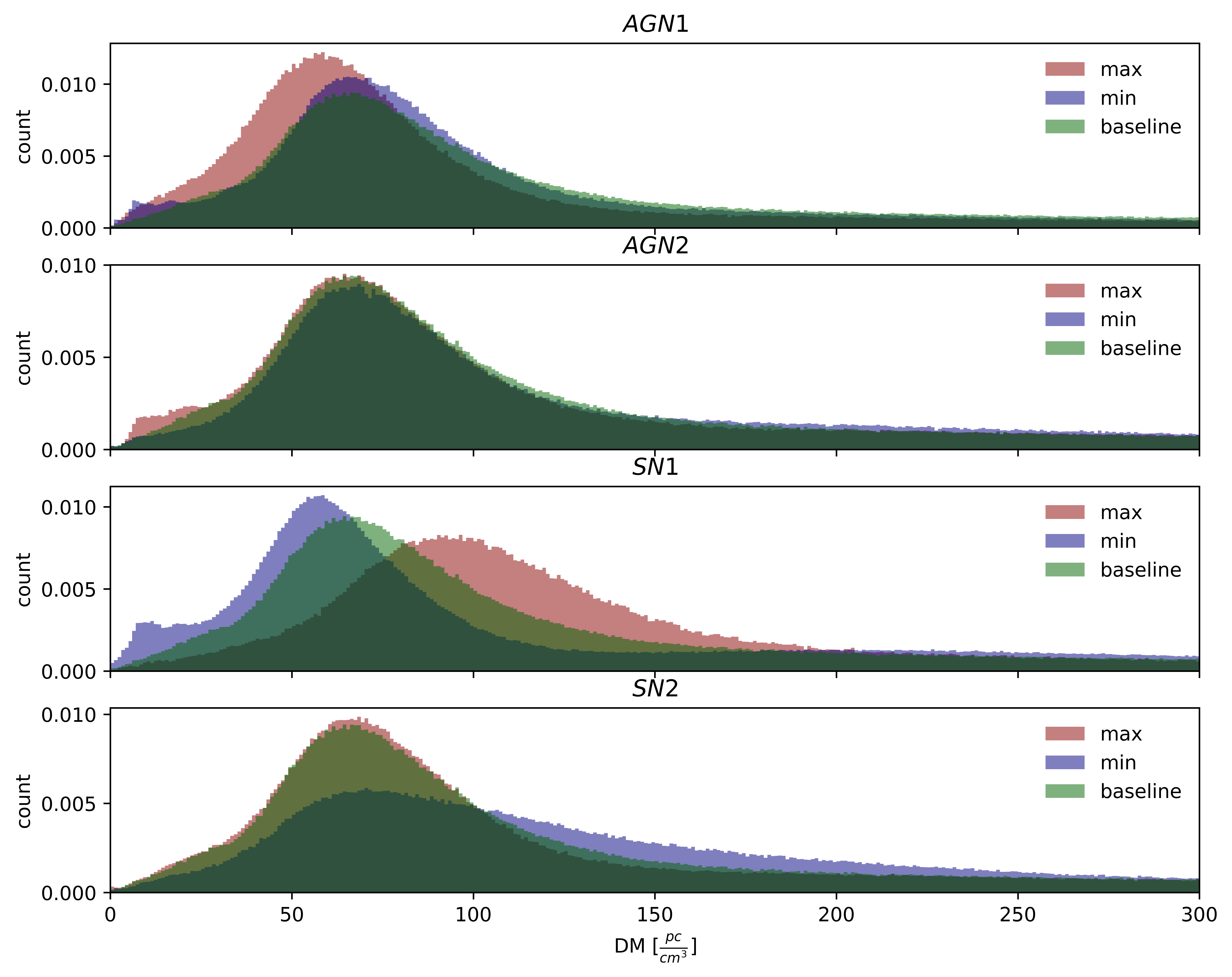}
    \caption[Effect of feedback on $\mathrm{DM}_{\mathrm{host}}$: Min. and max. parameter variation IllustrisTNG]{Dependency of the different feedback parameters in IllustrisTNG on the distribution of $\mathrm{DM}_{\mathrm{host}}$ at redshift $z=0$. The three different histograms show the maximum, minimum and baseline values for each of the parameters respectively.}
    \label{fig:feedback_min_max}
    \end{figure*}

\subsection{Host dispersion measure and halo properties}

     Our investigation into the dependency of $\mathrm{DM}_{\mathrm{host}}$ on halo mass reveals a consistent trend of increasing $\mathrm{DM}_{\mathrm{host}}$ with heavier halos across all three suites, \Cref{fig:m_halo_vs_DM}. This behaviour aligns with expectations, given that $\mathrm{DM}_{\mathrm{host}}$ is sensitive to the integrated electron density along the line of sight $\int \mathrm{d} l \,n_e$, which increases with the mass content within a halo. However, our findings also hint at a potential decline in $\mathrm{DM}_{\mathrm{host}}$ when analyzing the highest halo mass range considered. This decline may be attributed to the limited number of high-mass halos in the CAMELS simulations, a limitation resulting from the simulation's small box size, as highlighted by \citet{Villaescusa_Navarro_2021}, see also Appendix \ref{app:halos}.
     Furthermore, we study the host contribution as a function of SFR in \Cref{fig:SFR_vs_DM}, complementing the results from \Cref{fig:redshift_dependency} where we discussed the dependency of $\mathrm{DM}_{\mathrm{host}}$ on redshift $z$ .

    The figure demonstrates that stronger star formation amplifies the contribution of $\mathrm{DM}_{\mathrm{host}}$, confirming previous suggestions by \cite{Niu_2022} based on observations. Furthermore, we observed a similar correlation between the halo's metallicity and $\mathrm{DM}_{\mathrm{host}}$ (see Appendix \ref{sec:metallicity}). However, this correlation is anticipated, given that metals enhance gas condensation, leading to increased star formation. Simultaneously, stellar feedback influences the metallicity by releasing heavier elements into the star's environment. In addition, our examination of the redshift dependency of $\mathrm{DM}_{\mathrm{host}}$ reveals a slight increase in the maximum probability value of $\mathrm{DM}_{\mathrm{host}}$ when transitioning from higher to lower redshifts within the considered interval $z \in [0.00, 1.05]$. This observation aligns with expectations, considering that the halos under consideration are anticipated to grow towards $z = 0$. The evolving trend of $\mathrm{DM}_{\mathrm{host}}$ with redshift is also illustrated in figure \ref{fig:redshift_dependency}. The plot showcases the total $\mathrm{DM}_{\mathrm{host}}$ distribution for both placement modes for the IllustrisTNG suite across four redshift bins. It is noteworthy that the observed redshift evolution persists even when scrutinizing individual halo mass bins. Nevertheless, it is crucial to acknowledge once more the limitations of this study, primarily stemming from the small box size of the CAMELS simulations, which imposes constraints on the number of high-mass halos available for analysis. \\

    Since we find that the star formation rate is the halo property most strongly correlated with $\mathrm{DM}_\mathrm{host}$ and largely explains the other scalings, we fit a model for $\mathrm{DM}_{\mathrm{host}}(\mathrm{SFR})$ for the IllustrisTNG suite to further refine our understanding of star formation rate (SFR) dependency of $\mathrm{DM}_{\mathrm{host}}$ and use our model for future observations to achieve a better determination of the cosmological parameters. Leveraging the suite's comprehensive data, our numerical simulation allowed us to build a model of the mean value of the host DM for the joined sample of screenshots with redshifts $z \in [0.00, 1.05]$, which is given by:
    \begin{equation}
    \mathrm{DM}_{\mathrm{host}}(\mathrm{SFR}) = \widehat{\mathrm{DM}} \cdot \mathrm{SFR}^{\alpha}\,,
    \end{equation}
 with the mean DM host $\widehat{\mathrm{DM}} = 196^{+32}_{-33} \; \mathrm{pc} \; \mathrm{cm}^{-3}$ and exponent $\alpha = 0.23^{+0.05}_{-0.07}$.

Additionally, we attempt to fit the distributions themselves by different functional forms (see Appendix \ref{app:fitting} for details). The most commonly adopted functional form for the host contribution is a log-normal contribution. We find that the log-normal distribution underpredicts the long tail of the host contribution distribution. Furthermore, we find that a Burr distribution 
is a better fit. In practical scenarios, however, this difference will be most likely irrelevant.

    \subsection{Feedback effects on host dispersion}
    \label{sec:feedbackdep}

    \noindent For the concluding segment of the results section, we conducted a qualitative investigation into the feedback dependency of $\mathrm{DM}_{\mathrm{host}}$. Our analysis revealed that, among the four available parameters, only $A_{\mathrm{AGN1}}$ and $A_{\mathrm{SN1}}$ significantly influence the $\mathrm{DM}_{\mathrm{host}}$ distribution. This is shown in \Cref{fig:feedback_min_max}, which depicts the total $\mathrm{DM}{\mathrm{host}}$ distributions for all four feedback modes in their maximal and minimal settings as well as the fiducial parameter scenario for the IllustrisTNG suite at $z=0$. The graph shows that a high value of $A_{\mathrm{AGN1}}$ leads to a shift in the distribution towards lower values, likely due to a significant redistribution of matter and free electrons outside of the halo induced by strong black hole feedback. Additionally, the graph demonstrates that $A_{\mathrm{AGN2}}$ and $A_{\mathrm{SN2}}$ have a mild influence on the statistic, while $A_{\mathrm{SN1}}$ emerges as the most relevant factor impacting the $\mathrm{DM}_{\mathrm{host}}$ distribution. An increase in $A_{\mathrm{SN1}}$ results in a higher MP value of $\mathrm{DM}_{\mathrm{host}}$, likely attributable to redistribution processes similar to those triggered by $A_{\mathrm{AGN1}}$, but with free electrons remaining within the halo. It is important to note that the findings for the SIMBA suite are similar to those for IllustrisTNG, albeit with a slightly less pronounced effect of $A_{\mathrm{AGN1}}$ and $A_{\mathrm{SN2}}$ on $\mathrm{DM}_{\mathrm{host}}$. However, the feedback effects are considerably less pronounced and most prominent for the $A_{\mathrm{AGN1}}$ parameter in the ASTRID simulations. Moreover, ASTRID's $\mathrm{DM}_{\mathrm{host}}$ distributions for minimal feedback parameter values exhibit a second peak when using the FRB placement mode following the total stellar mass. This phenomenon may be attributed to the redistribution of free electrons towards the halo's centre due to weaker feedback. Consequently, FRBs originating from sources in the outer halo regions experience less dispersion for specific directions. However, it is important to acknowledge potential limitations in our findings due to the use of the CAMELS simulations. The relatively small box size and resulting limited number of large halos in these simulations could significantly alter distributions, especially given the strong association of supernovae (SN) and black holes with high-mass halos. Additionally, AGN and SN feedback mechanisms influence each other, and changing one parameter may indirectly affect others, as suggested by \cite{Nicola_2022}.

\section{Conclusion}
\label{sec:conclusion}
We perform a detailed study of the evolution of the dispersion measure contributions from FRB host galaxies with redshift, for different feedback models and different simulation codes with the CAMELS simulation suite. We explore the effect of different FRB placement modes (either according to the total stellar mass, or to the star formation rate) representing different possible formation channels, but find negligible differences in the inferred dispersion measure.

The by far largest differences occur between the different baryonic feedback models of the Illustris, SIMBA and ASTRID simulation suites. Especially the comparably weak feedback from ASTRID leads to very large values of the host galaxy dispersion measure. Current data is not yet sufficiently accurate to distinguish between the models, but a modest increase in localised FRBs could potentially differentiate between them. This opens the way to calibrate baryonic feedback models directly from measured host galaxy dispersion once larger FRB samples are available in the near future and could provide an important step in order to model the non-linear power spectrum at scales crucial for ongoing stage-IV cosmological surveys. The fiducial feedback model of the ASTRID simulations is already under slight tension with current measurements of FRB host galaxy dispersions, but a sample of a few hundred localised events should be sufficient to inform feedback implementations in cosmological simulations. FRBs can here be complementary to other observations of the baryon distribution via e.g. the Sunyaev-Zel'dovich effect \citep{Troster:2021gsz, Pandey:2023wqp}

We find the evolution of the host dispersion with halo mass and redshift, but those are mostly accounted for by varying star formation rates of halos over cosmic time. We find an approximate power law relation between $\mathrm{DM}_\mathrm{host}$ and the star formation rate, which can be used as a prior for cosmological applications of localised FRBs with known spectral host properties, therefore improving cosmological parameter constraints in studies such as \citep{Hagstotz_2022, James_2022}.

\begin{figure}
    \centering
    \includegraphics[width = 0.47\textwidth]{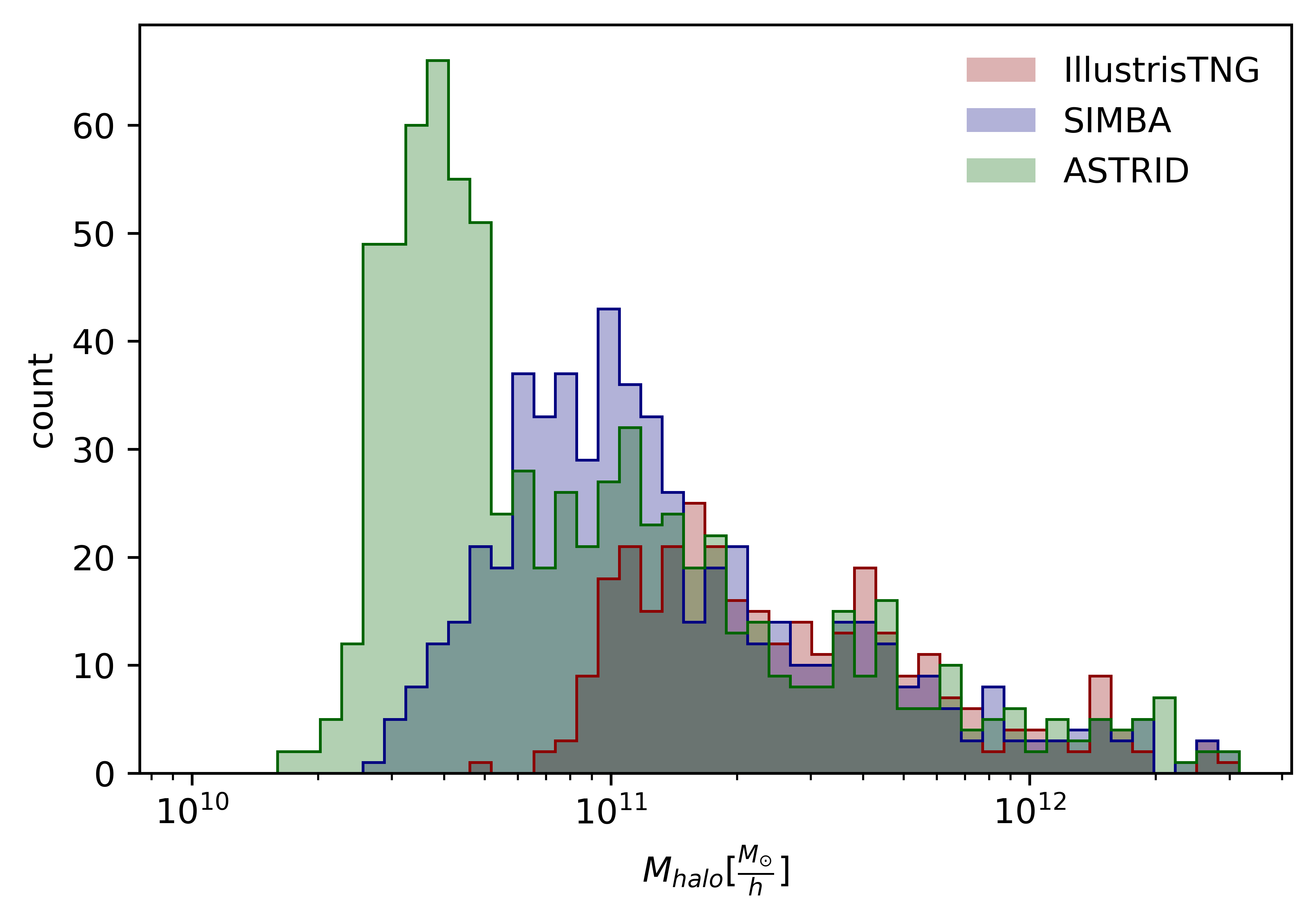}
    \caption{Histogram of the number of halos as a function of halo mass in the simulation suites used in CAMELS.}
    \label{fig:halonumber}
\end{figure}

$\bf{Note:}$ As this paper was about to be submitted, the authors of \cite{Medlock_2024} published a similar, independent analysis using the CAMELS simulation suite and came to the same results concerning the strong influence of different feedback implementation on the dispersion measure. While \cite{Medlock_2024} focus on the variance of the $\mathrm{DM}(z)$ relation and its variation with feedback strength, we provide the distribution of the halo dispersion measure averaged over all lines of sight as an input for FRB analyses.

\section*{Acknowledgements}
We thank the CAMELS team for sharing their simulation suite openly with the community. We thank the authors of \cite{Macquart:2020lln} for making their numerical tools for FRB-related calculations publicly available. We thank Klaus Dolag, Andrina Nicola, Kai Lehmann and Francisco Villaescusa-Navarro for valuable discussions about the simulations, and we thank Nico Hamaus for help with density field estimators.
SH was supported by the Excellence Cluster ORIGINS which is funded by the Deutsche Forschungsgemeinschaft (DFG, German Research Foundation) under Germany’s Excellence Strategy - EXC-2094 - 390783311. RR is supported by the European Research Council (Grant No. 770935).

\section*{Data Availability}

The CAMELS simulations are publicly available.



\bibliographystyle{mnras}
\bibliography{bibliography}

\begin{thebibliography}{}
\makeatletter
\relax
\def\mn@urlcharsother{\let\do\@makeother \do\$\do\&\do\#\do\^\do\_\do\%\do\~}
\def\mn@doi{\begingroup\mn@urlcharsother \@ifnextchar [ {\mn@doi@}
  {\mn@doi@[]}}
\def\mn@doi@[#1]#2{\def\@tempa{#1}\ifx\@tempa\@empty \href
  {http://dx.doi.org/#2} {doi:#2}\else \href {http://dx.doi.org/#2} {#1}\fi
  \endgroup}
\def\mn@eprint#1#2{\mn@eprint@#1:#2::\@nil}
\def\mn@eprint@arXiv#1{\href {http://arxiv.org/abs/#1} {{\tt arXiv:#1}}}
\def\mn@eprint@dblp#1{\href {http://dblp.uni-trier.de/rec/bibtex/#1.xml}
  {dblp:#1}}
\def\mn@eprint@#1:#2:#3:#4\@nil{\def\@tempa {#1}\def\@tempb {#2}\def\@tempc
  {#3}\ifx \@tempc \@empty \let \@tempc \@tempb \let \@tempb \@tempa \fi \ifx
  \@tempb \@empty \def\@tempb {arXiv}\fi \@ifundefined
  {mn@eprint@\@tempb}{\@tempb:\@tempc}{\expandafter \expandafter \csname
  mn@eprint@\@tempb\endcsname \expandafter{\@tempc}}}

\bibitem[\protect\citeauthoryear{{Alonso}}{{Alonso}}{2021}]{alonso_linear_2021}
{Alonso} D.,  2021, \mn@doi [\prd] {10.1103/PhysRevD.103.123544}, \href
  {https://ui.adsabs.harvard.edu/abs/2021PhRvD.103l3544A} {103, 123544}

\bibitem[\protect\citeauthoryear{Aver, Olive  \& Skillman}{Aver
  et~al.}{2015}]{Aver:2015iza}
Aver E.,  Olive K.~A.,   Skillman E.~D.,  2015, \mn@doi [JCAP]
  {10.1088/1475-7516/2015/07/011}, 07, 011

\bibitem[\protect\citeauthoryear{Bhattacharya, Kumar  \& Linder}{Bhattacharya
  et~al.}{2020}]{bhattacharya_fast_2020}
Bhattacharya M.,  Kumar P.,   Linder E.~V.,  2020, arXiv:2010.14530

\bibitem[\protect\citeauthoryear{{Bird}, {Ni}, {Di Matteo}, {Croft}, {Feng}  \&
  {Chen}}{{Bird} et~al.}{2022}]{bird_astrid_2022}
{Bird} S.,  {Ni} Y.,  {Di Matteo} T.,  {Croft} R.,  {Feng} Y.,   {Chen} N.,
  2022, \mn@doi [\mnras] {10.1093/mnras/stac648}, \href
  {https://ui.adsabs.harvard.edu/abs/2022MNRAS.512.3703B} {512, 3703}

\bibitem[\protect\citeauthoryear{Bochenek, Ravi, Belov, Hallinan, Kocz,
  Kulkarni  \& McKenna}{Bochenek et~al.}{2020}]{Bochenek_2020}
Bochenek C.~D.,  Ravi V.,  Belov K.~V.,  Hallinan G.,  Kocz J.,  Kulkarni
  S.~R.,   McKenna D.~L.,  2020, \mn@doi [Nature] {10.1038/s41586-020-2872-x},
  587, 59

\bibitem[\protect\citeauthoryear{Champion et~al.,}{Champion
  et~al.}{2016}]{champion_five_2016}
Champion D.~J.,  et~al., 2016, \mn@doi [Monthly Notices of the Royal
  Astronomical Society] {10.1093/mnrasl/slw069}, 460, L30

\bibitem[\protect\citeauthoryear{Chatterjee et~al.,}{Chatterjee
  et~al.}{2017}]{chatterjee_direct_2017}
Chatterjee S.,  et~al., 2017, \mn@doi [Nature] {10.1038/nature20797}, 541, 58

\bibitem[\protect\citeauthoryear{Connor, Sievers  \& Pen}{Connor
  et~al.}{2016}]{connor_non-cosmological_2016}
Connor L.,  Sievers J.,   Pen U.-L.,  2016, \mn@doi [Monthly Notices of the
  Royal Astronomical Society] {10.1093/mnrasl/slv124}, 458, L19

\bibitem[\protect\citeauthoryear{Cordes \& Lazio}{Cordes \&
  Lazio}{2003}]{cordes2003ne2001}
Cordes J.~M.,  Lazio T. J.~W.,  2003, NE2001. II. Using Radio Propagation Data
  to Construct a Model for the Galactic Distribution of Free Electrons
  (\mn@eprint {arXiv} {astro-ph/0301598})

\bibitem[\protect\citeauthoryear{{Dav{\'e}}, {Angl{\'e}s-Alc{\'a}zar},
  {Narayanan}, {Li}, {Rafieferantsoa}  \& {Appleby}}{{Dav{\'e}}
  et~al.}{2019}]{dave_2019_simba}
{Dav{\'e}} R.,  {Angl{\'e}s-Alc{\'a}zar} D.,  {Narayanan} D.,  {Li} Q.,
  {Rafieferantsoa} M.~H.,   {Appleby} S.,  2019, \mn@doi [\mnras]
  {10.1093/mnras/stz937}, \href
  {https://ui.adsabs.harvard.edu/abs/2019MNRAS.486.2827D} {486, 2827}

\bibitem[\protect\citeauthoryear{Dolag, Borgani, Murante  \& Springel}{Dolag
  et~al.}{2009}]{Dolag_2009}
Dolag K.,  Borgani S.,  Murante G.,   Springel V.,  2009, \mn@doi [Monthly
  Notices of the Royal Astronomical Society]
  {10.1111/j.1365-2966.2009.15034.x}, 399, 497

\bibitem[\protect\citeauthoryear{Hagstotz, Reischke  \& Lilow}{Hagstotz
  et~al.}{2022}]{Hagstotz_2022}
Hagstotz S.,  Reischke R.,   Lilow R.,  2022, \mn@doi [Monthly Notices of the
  Royal Astronomical Society] {10.1093/mnras/stac077}, 511, 662

\bibitem[\protect\citeauthoryear{{Hopkins}}{{Hopkins}}{2015}]{hopkins_gizmo_2015}
{Hopkins} P.~F.,  2015, \mn@doi [\mnras] {10.1093/mnras/stv195}, \href
  {https://ui.adsabs.harvard.edu/abs/2015MNRAS.450...53H} {450, 53}

\bibitem[\protect\citeauthoryear{James et~al.,}{James
  et~al.}{2022}]{James_2022}
James C.~W.,  et~al., 2022, \mn@doi [Monthly Notices of the Royal Astronomical
  Society] {10.1093/mnras/stac2524}, 516, 4862

\bibitem[\protect\citeauthoryear{{Khrykin} et~al.,}{{Khrykin}
  et~al.}{2024}]{flimflam_2024}
{Khrykin} I.~S.,  et~al., 2024, \mn@doi [arXiv e-prints]
  {10.48550/arXiv.2402.00505}, \href
  {https://ui.adsabs.harvard.edu/abs/2024arXiv240200505K} {p. arXiv:2402.00505}

\bibitem[\protect\citeauthoryear{{Kirsten} et~al.,}{{Kirsten}
  et~al.}{2022}]{Kirsten:2021llv}
{Kirsten} F.,  et~al., 2022, \mn@doi [\nat] {10.1038/s41586-021-04354-w}, \href
  {https://ui.adsabs.harvard.edu/abs/2022Natur.602..585K} {602, 585}

\bibitem[\protect\citeauthoryear{Liu, Romero, Liu  \& Li}{Liu
  et~al.}{2016}]{Liu_2016}
Liu T.,  Romero G.~E.,  Liu M.-L.,   Li A.,  2016, \mn@doi [The Astrophysical
  Journal] {10.3847/0004-637x/826/1/82}, 826, 82

\bibitem[\protect\citeauthoryear{Macquart et~al.}{Macquart
  et~al.}{2020}]{Macquart:2020lln}
Macquart J.~P.,  et~al., 2020, \mn@doi [Nature] {10.1038/s41586-020-2300-2},
  581, 391

\bibitem[\protect\citeauthoryear{Madau \& Dickinson}{Madau \&
  Dickinson}{2014}]{Madau:2014bja}
Madau P.,  Dickinson M.,  2014, \mn@doi [Ann. Rev. Astron. Astrophys.]
  {10.1146/annurev-astro-081811-125615}, 52, 415

\bibitem[\protect\citeauthoryear{Masui \& Sigurdson}{Masui \&
  Sigurdson}{2015}]{masui_dispersion_2015}
Masui K.~W.,  Sigurdson K.,  2015, \mn@doi [Physical Review Letters]
  {10.1103/PhysRevLett.115.121301}, 115, 121301

\bibitem[\protect\citeauthoryear{{Medlock}, {Nagai}, {Singh}, {Oppenheimer},
  {Angl{\'e}s Alc{\'a}zar}  \& {Villaescusa-Navarro}}{{Medlock}
  et~al.}{2024}]{Medlock_2024}
{Medlock} I.,  {Nagai} D.,  {Singh} P.,  {Oppenheimer} B.,  {Angl{\'e}s
  Alc{\'a}zar} D.,   {Villaescusa-Navarro} F.,  2024, \mn@doi [arXiv e-prints]
  {10.48550/arXiv.2403.02313}, \href
  {https://ui.adsabs.harvard.edu/abs/2024arXiv240302313M} {p. arXiv:2403.02313}

\bibitem[\protect\citeauthoryear{Mo, Zhu, Wang, Tang  \& Feng}{Mo
  et~al.}{2022}]{Mo_2022}
Mo J.-F.,  Zhu W.,  Wang Y.,  Tang L.,   Feng L.-L.,  2022, \mn@doi [Monthly
  Notices of the Royal Astronomical Society] {10.1093/mnras/stac3104}, 518, 539

\bibitem[\protect\citeauthoryear{{Ni} et~al.,}{{Ni}
  et~al.}{2022}]{ni_astrid_2022}
{Ni} Y.,  et~al., 2022, \mn@doi [\mnras] {10.1093/mnras/stac351}, \href
  {https://ui.adsabs.harvard.edu/abs/2022MNRAS.513..670N} {513, 670}

\bibitem[\protect\citeauthoryear{Ni et~al.,}{Ni et~al.}{2023}]{ni2023camels}
Ni Y.,  et~al., 2023, The CAMELS project: Expanding the galaxy formation model
  space with new ASTRID and 28-parameter TNG and SIMBA suites (\mn@eprint
  {arXiv} {2304.02096})

\bibitem[\protect\citeauthoryear{Nicola et~al.,}{Nicola
  et~al.}{2022}]{Nicola_2022}
Nicola A.,  et~al., 2022, \mn@doi [Journal of Cosmology and Astroparticle
  Physics] {10.1088/1475-7516/2022/04/046}, 2022, 046

\bibitem[\protect\citeauthoryear{Niu et~al.,}{Niu et~al.}{2022}]{Niu_2022}
Niu C.-H.,  et~al., 2022, \mn@doi [Nature] {10.1038/s41586-022-04755-5}, 606,
  873

\bibitem[\protect\citeauthoryear{Pandey et~al.,}{Pandey
  et~al.}{2023}]{Pandey:2023wqp}
Pandey S.,  et~al., 2023, \mn@doi [Mon. Not. Roy. Astron. Soc.]
  {10.1093/mnras/stad2268}, 525, 1779

\bibitem[\protect\citeauthoryear{Petroff et~al.,}{Petroff
  et~al.}{2015}]{petroff_real-time_2015}
Petroff E.,  et~al., 2015, \mn@doi [MNRAS] {10.1093/mnras/stu2419}, 447, 246

\bibitem[\protect\citeauthoryear{{Pillepich} et~al.,}{{Pillepich}
  et~al.}{2018}]{pillepic_tng_2018}
{Pillepich} A.,  et~al., 2018, \mn@doi [\mnras] {10.1093/mnras/stx2656}, \href
  {https://ui.adsabs.harvard.edu/abs/2018MNRAS.473.4077P} {473, 4077}

\bibitem[\protect\citeauthoryear{{Rafiei-Ravandi}, Smith  \&
  Masui}{{Rafiei-Ravandi} et~al.}{2020}]{rafiei-ravandi_characterizing_2020}
{Rafiei-Ravandi} M.,  Smith K.~M.,   Masui K.~W.,  2020, \mn@doi [Physical
  Review D] {10.1103/PhysRevD.102.023528}, 102, 023528

\bibitem[\protect\citeauthoryear{{Reischke} \& {Hagstotz}}{{Reischke} \&
  {Hagstotz}}{2023a}]{reischke_consistent_2023}
{Reischke} R.,  {Hagstotz} S.,  2023a, \mn@doi [\mnras]
  {10.1093/mnras/stad1866}, \href
  {https://ui.adsabs.harvard.edu/abs/2023MNRAS.523.6264R} {523, 6264}

\bibitem[\protect\citeauthoryear{Reischke \& Hagstotz}{Reischke \&
  Hagstotz}{2023b}]{reischke_cosmological_2023}
Reischke R.,  Hagstotz S.,  2023b, \mn@doi [Monthly Notices of the Royal
  Astronomical Society] {10.1093/mnras/stad1645}, 524, 2237

\bibitem[\protect\citeauthoryear{Reischke, Hagstotz  \& Lilow}{Reischke
  et~al.}{2021a}]{Reischke:2021euf}
Reischke R.,  Hagstotz S.,   Lilow R.,  2021a, arXiv:2102.11554

\bibitem[\protect\citeauthoryear{Reischke, Hagstotz  \& Lilow}{Reischke
  et~al.}{2021b}]{Reischke:2020cgd}
Reischke R.,  Hagstotz S.,   Lilow R.,  2021b, \mn@doi [Phys. Rev. D]
  {10.1103/PhysRevD.103.023517}, 103, 023517

\bibitem[\protect\citeauthoryear{{Reischke}, {Neumann}, {Bertmann}, {Hagstotz}
  \& {Hildebrandt}}{{Reischke} et~al.}{2023}]{reischke_feedback_2023}
{Reischke} R.,  {Neumann} D.,  {Bertmann} K.~A.,  {Hagstotz} S.,
  {Hildebrandt} H.,  2023, \mn@doi [arXiv e-prints]
  {10.48550/arXiv.2309.09766}, \href
  {https://ui.adsabs.harvard.edu/abs/2023arXiv230909766R} {p. arXiv:2309.09766}

\bibitem[\protect\citeauthoryear{Shirasaki, Kashiyama  \& Yoshida}{Shirasaki
  et~al.}{2017}]{shirasaki_large-scale_2017}
Shirasaki M.,  Kashiyama K.,   Yoshida N.,  2017, \mn@doi [Physical Review D]
  {10.1103/PhysRevD.95.083012}, 95, 083012

\bibitem[\protect\citeauthoryear{Springel, White, Tormen  \&
  Kauffmann}{Springel et~al.}{2001}]{Springel:01}
Springel V.,  White S. D.~M.,  Tormen G.,   Kauffmann G.,  2001, \mn@doi
  [Monthly Notices of the Royal Astronomical Society]
  {10.1046/j.1365-8711.2001.04912.x}, 328, 726

\bibitem[\protect\citeauthoryear{Takahashi, Ioka, Mori  \& Funahashi}{Takahashi
  et~al.}{2021}]{takahashi_statistical_2021}
Takahashi R.,  Ioka K.,  Mori A.,   Funahashi K.,  2021, \mn@doi [Monthly
  Notices of the Royal Astronomical Society] {10.1093/mnras/stab170}, 502, 2615

\bibitem[\protect\citeauthoryear{Thornton et~al.,}{Thornton
  et~al.}{2013a}]{thornton_population_2013}
Thornton D.,  et~al., 2013a, \mn@doi [Science] {10.1126/science.1236789}, 341,
  53

\bibitem[\protect\citeauthoryear{Thornton et~al.,}{Thornton
  et~al.}{2013b}]{Thornton_2013}
Thornton D.,  et~al., 2013b, \mn@doi [Science] {10.1126/science.1236789}, 341,
  53

\bibitem[\protect\citeauthoryear{Tr\"oster et~al.}{Tr\"oster
  et~al.}{2022}]{Troster:2021gsz}
Tr\"oster T.,  et~al., 2022, \mn@doi [Astron. Astrophys.]
  {10.1051/0004-6361/202142197}, 660, A27

\bibitem[\protect\citeauthoryear{Villaescusa-Navarro
  et~al.,}{Villaescusa-Navarro et~al.}{2021}]{Villaescusa_Navarro_2021}
Villaescusa-Navarro F.,  et~al., 2021, \mn@doi [The Astrophysical Journal]
  {10.3847/1538-4357/abf7ba}, 915, 71

\bibitem[\protect\citeauthoryear{Villaescusa-Navarro
  et~al.,}{Villaescusa-Navarro et~al.}{2023}]{Villaescusa_Navarro_2023}
Villaescusa-Navarro F.,  et~al., 2023, \mn@doi [The Astrophysical Journal
  Supplement Series] {10.3847/1538-4365/acbf47}, 265, 54

\bibitem[\protect\citeauthoryear{{Weinberger} et~al.,}{{Weinberger}
  et~al.}{2017}]{weinberger_tng_2017}
{Weinberger} R.,  et~al., 2017, \mn@doi [\mnras] {10.1093/mnras/stw2944}, \href
  {https://ui.adsabs.harvard.edu/abs/2017MNRAS.465.3291W} {465, 3291}

\bibitem[\protect\citeauthoryear{Wu, Yu  \& Wang}{Wu et~al.}{2020}]{Wu:2020jmx}
Wu Q.,  Yu H.,   Wang F.~Y.,  2020, \mn@doi [Astrophys. J.]
  {10.3847/1538-4357/ab88d2}, 895, 33

\bibitem[\protect\citeauthoryear{{Yao}, {Manchester}  \& {Wang}}{{Yao}
  et~al.}{2017}]{YMW16}
{Yao} J.~M.,  {Manchester} R.~N.,   {Wang} N.,  2017, \mn@doi [\apj]
  {10.3847/1538-4357/835/1/29}, \href
  {https://ui.adsabs.harvard.edu/abs/2017ApJ...835...29Y} {835, 29}

\bibitem[\protect\citeauthoryear{Zhou, Li, Wang, Fan  \& Wei}{Zhou
  et~al.}{2014}]{zhou_fast_2014}
Zhou B.,  Li X.,  Wang T.,  Fan Y.-Z.,   Wei D.-M.,  2014, \mn@doi [Phys. Rev.
  D] {10.1103/PhysRevD.89.107303}, 89, 107303

\bibitem[\protect\citeauthoryear{for~the TNG~Collaboration.}{for~the
  TNG~Collaboration.}{2023}]{TNG}
for~the TNG~Collaboration. D.~N.,  2023, Public Data Access Overview / Data
  Specifications, \url
  {https://www.tng-project.org/data/docs/specifications/#sec1a}

\makeatother
\end{thebibliography}

\appendix

\section{Number of halos in the simulation}\label{app:halos}
For completeness, we show the number of halos as a function of halo mass found in the different simulation suites. From \Cref{fig:halonumber} one can clearly see that there is only a handful of halos with $M_\mathrm{halo} > 10^{12}\, h^{-1}M_\odot $ due to the limited CAMELS box size.

\section{Fitting the host contribution}
\begin{figure}
    \centering
    \includegraphics[width = .47\textwidth]{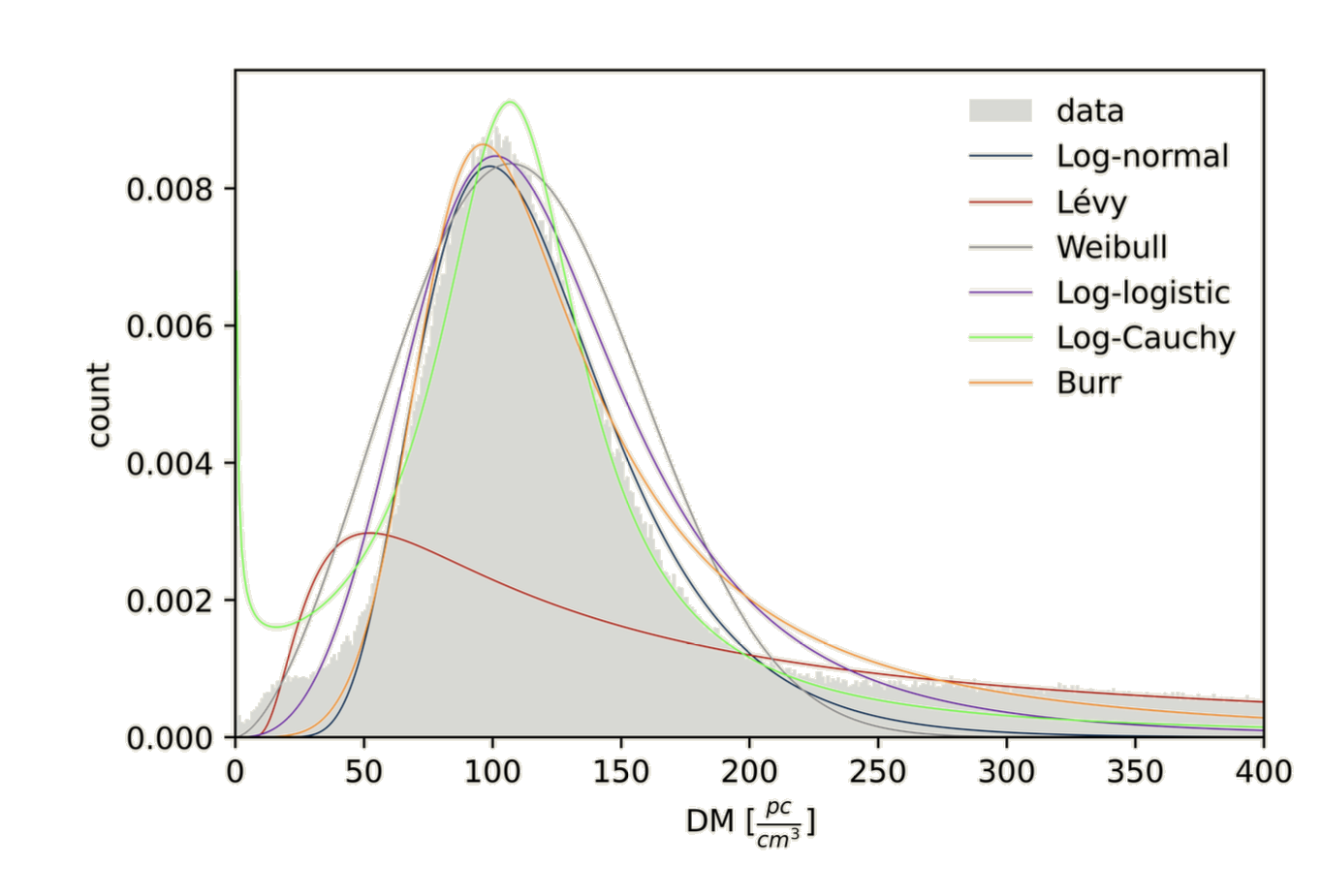}
    \caption{Dispersion measure host contribution for Illustris-TNG with different functional fits (colour-coded).}
    \label{fig:fitting_distribution}
\end{figure}
\label{app:fitting}
In \Cref{fig:fitting_distribution} we show fits to the $\mathrm{DM}_\mathrm{host}$ contribution using different functional forms. In particular, we use the commonly adopted log-normal distribution, with $x = \mathrm{DM}_\mathrm{host}$:
\begin{equation}
    p(x) = \frac{1}{x\sigma\sqrt{2\pi}}\exp\left(-\frac{(\ln x - \mu)}{2\sigma^2}\right)\,,
\end{equation}
with the logarithmic mean, $\mu$, and variance, $\sigma$.
The L\'{e}vi distribution
\begin{equation}
    p(x) = \sqrt{\frac{C}{2\pi}}\frac{\exp\left(-\frac{C}{2(x-\mu)}\right)}{(x-\mu)^3/2}\,,
\end{equation}
with the location parameter $\mu$ and the scale parameter $C$. Lastly, a generalised logarithmic logistic distribution, the Burr distribution:
\begin{equation}
    p(x) = Ck \frac{x^{C-1}}{(1+x^C)^{k+1}}\;.
\end{equation}
We also consider other forms of log-logistic functions as well as the Weibull and log-Cauchy distributions. We generally find that the L\'{e}vy distribution can describe the long tail of the distributions best. It fails, however, to describe the core accurately and therefore is overall not a good description of the host distribution. Generally, the Burr distribution does the best job of fitting both the core as well as the tail. The difference to the commonly used log-normal distribution is, however, marginal since the bulk of the events are well captured by the core. 

\section{$\mathrm{DM}_\mathrm{host}$ as a function of metallicity}
\label{sec:metallicity}
\begin{figure}
    \centering
    \includegraphics[width = 0.47\textwidth]{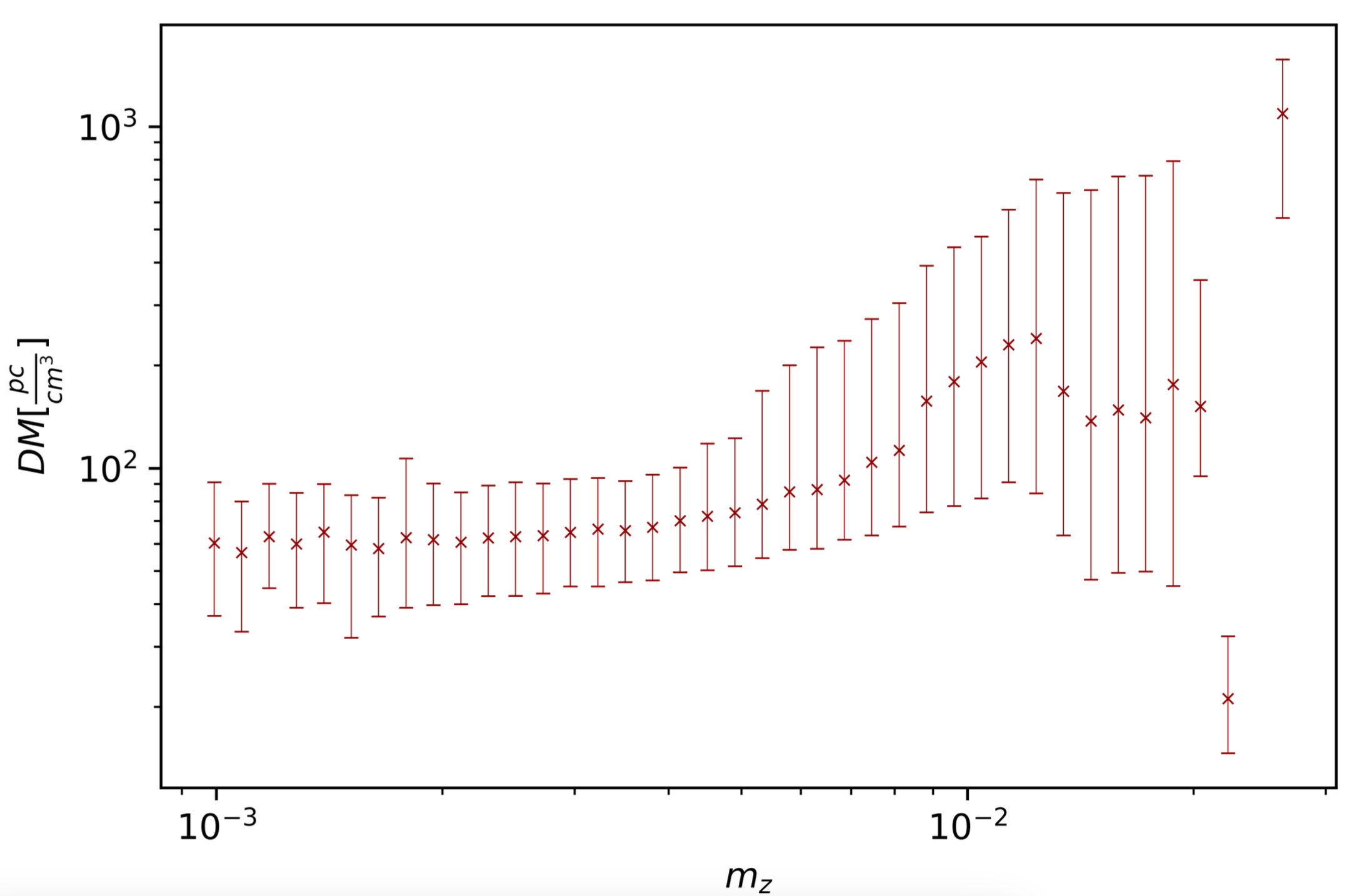}
    \caption{DM host contribution as a function of metallicity (compare \Cref{fig:m_halo_vs_DM}).}
    \label{fig:metallicity}
\end{figure}

\Cref{fig:metallicity} shows the connection between halo's metallicity and the $\mathrm{DM}_\mathrm{host}$ contribution. This trend is expected as metals enhance gas condensation thus increasing star formation while also stellar feedback influences the metallicity itself by releasing heavy elements into the ambient structures.
\end{document}